\documentclass[remotesensing,article,accept,pdftex,moreauthors]{Definitions/mdpi} 

\firstpage{1} 
\pubvolume{1}
\issuenum{1}
\articlenumber{0}
\pubyear{2024}
\copyrightyear{2024}
\externaleditor{Academic Editors: Roberto Orosei and Giancarlo Bellucci}
\datereceived{28 February 2024 } 
\daterevised{23 March 2024 } 
\dateaccepted{24 March 2024 } 
\datepublished{ } 
\hreflink{https://doi.org/} 

\Title{Dynamical Model of Rotation and Orbital Coupling for Deimos}

\TitleCitation{Dynamical Model of Rotation and Orbital Coupling for Deimos}

\Author{Kai Huang 
 $^{1,2}$, Lijun Zhang $^{1,2}$, Yongzhang Yang $^{1,3}$, Mao Ye $^{4}$ and Yuqiang Li $^{1,5,}$*}

\AuthorNames{Kai Huang, Lijun Zhang and Yongzhang Yang}

\AuthorCitation{Huang, K.; 
 Zhang, L.; Yang, Y.; Ye, M.; Li, Y.}

\address{%
$^{1}$ \quad Yunnan Observatories, Chinese Academy of Sciences, Kunming 650216, China; huangkai@ynao.ac.cn (K.H.); zhanglijun@ynao.ac.cn (L.Z.); yang.yongzhang@ynao.ac.cn (Y.Y.)\\
$^{2}$ \quad University of Chinese Academy of Sciences, Beijing
100049, China\\
$^{3}$ \quad Project Supported by MOE Key Laboratory of TianQin Project, Sun Yat-Sen University, Zhuhai 519082, China\\
$^{4}$ \quad State Key Laboratory of Information Engineering in Surveying, Mapping and Remote Sensing, \mbox{Wuhan University,} Wuhan 430079, China; mye@whu.edu.cn\\
$^{5}$ \quad Key Laboratory of Space Object and Debris Observation, PMO, CAS, Nanjing 210023, China}
\corres{Correspondence: lyq@ynao.ac.cn} 

\abstract{This paper introduces a novel dynamical model, building upon the existing dynamical model for Deimos in the current numerical ephemerides, which only encompasses the simple libration effects of Deimos. The study comprehensively incorporates the rotational dynamics of Deimos influenced by the torque exerted by the major celestial bodies (Mars, the Sun) in the solar system within the inertial space. Consequently, a full dynamical model is formulated to account for the complete coupling between the rotation and orbit of Deimos. Simultaneously, employing precision orbit determination methods used for artificial satellites, we develop an adjustment model for fitting data to the complete model. The 12-order Adams--Bashforth--Moulton (ABM) integration algorithm is employed to synchronously integrate the 12 state variables of the full model to obtain the orbit of Deimos.The difference in the orbits obtained by integrating the full model over a period of 10 years and those obtained by the simplified model is at the order of 10 km. After precise orbit determination, this difference decreases to below 100 m, so numerical simulation results indicate that the full dynamical model and adjustment model are stable and reliable. Simultaneously, the integration of the Deimos third-order gravity field in the full model over a 10-year period induces only meter-level positional changes. This suggests that when constructing the complete model, the utilization of a second-order gravity field alone is sufficient. Compared to the simple model, the polar axis of Deimos in the inertial space exhibits a more complex oscillation in the full model. Additionally, the full model calculates that the minimum moment of inertia principal axis of Phobos has an amplitude of approximately 0.5 degrees in the longitude direction and does not exceed 2 degrees in the latitude direction. This work further advances the current dynamical model for Deimos and establishes the foundational model for the generation of a new set of precise numerical ephemerides for Deimos.}


\begin{document}



\section{Introduction}

Human understanding of the Martian moons began with the discovery by Asaph Hall, a scientist at the United States Naval Observatory, in 1877. He identified the two moons of Mars and named them Phobos and Deimos, respectively. Researchers have conducted comprehensive studies on Phobos \cite{Kamada2024Modeling4B,Kuramoto2021MartianME}. Compared to Phobos, there is relatively limited publicly available research on Deimos. Therefore, this paper focuses on Deimos as the primary research target. The physical parameters of Deimos are shown in Table~\ref{tab2}~\cite{Rubincam1995TheGF}. The study of the Martian moons not only contributes to a deeper understanding of the formation and evolution of terrestrial planet systems but also provides crucial clues in researching the formation and evolution of the solar system.
Up to now, humans have launched multiple probes to Mars and conducted extensive exploration activities on its two moons. Approximately 90 years after the initial discovery of the two Martian moons, the first artificial satellite dedicated to Mars, Mariner 4, conducted a flyby of Mars in 1965~\cite{Anderson1965MarinerIM}. Subsequently, human exploration of Mars has not ceased, and spacecraft have been launched to Mars on multiple occasions. Mariners 6 and 7 also carried out flyby missions of Mars. In 1971, the first Mars orbiter, Mariner 9, was launched. It obtained 214 images of Phobos and Deimos during close flybys, marking the earliest spacecraft observations of the Martian moons.
In the late 1970s, the Soviet Union successively launched two orbiters, Viking 1 and Viking 2, to Mars. These spacecraft were equipped with landers. Among them, Mars Express, launched by the European Space Agency (ESA), acquired the most data on the Martian moons. Its impact on understanding the orbits of the Martian moons and the Martian ephemerides has been profound. Notably, data from the Pathfinder Mars lander played a crucial role in the study of Mars' rotation and gravity field. The gravity field model MRO120F from the Pathfinder mission is utilized in this research~\cite{Konopliv2016AnIJ}.
In recent years, both China and Japan have initiated their own Mars exploration missions~\cite{KJKZ202105002,Kuramoto2020MartianME,Croswell2023AnAN}. It is anticipated that an abundance of high-precision data will be instrumental in advancing the research on the dynamical models of Martian moons.
\begin{table}[H] 
	\caption{Major physics parameters of Deimos.\label{tab2}}
	\newcolumntype{C}{>{\centering\arraybackslash}X}
	\begin{tabularx}{\textwidth}{cCC}
		\toprule
		\textbf{Parameters}	&  \textbf{Symbol} & \textbf{Value}	\\
		\midrule
		Mass parameters & $\mathrm{GM}\left(\mathrm{m}^3 / \mathrm{s}^2\right)$ & $9.68 \times 10^{4}$	\\
		Radius & $\mathrm{R}(\mathrm{m})$		& 6250			\\
		Second-degree and order tesseral harmonic & $C_{22}$		& $4.773 \times 10^{-2}$		\\
		Second-degree zonal harmonic&$J_2$		& 0.108		\\
		The moments of inertia &$\mathrm{~A}$ & 0.338 \\
		 &$\mathrm{~B}$ & 0.461 \\
		 &$\mathrm{~C}$  & 0.508 \\
		\bottomrule
	\end{tabularx}
\end{table}
In various exploration activities of celestial bodies beyond Earth, investigating their orbital trajectories is a crucial scientific inquiry, holding significant academic importance for a profound understanding of the internal structure and physical parameters of these celestial entities. The data collected from ground-based observations to spacecraft missions provide the foundation for our research on the dynamical models of Martian moons.      
Multiple international institutions have conducted in-depth research in constructing dynamical models for Martian moons. With the improvement in the volume and precision of human exploration data for Martian moons, the dynamical models for these moons have undergone a developmental process from analytical models and semi-analytical models to numerical models.

In the early stages, dynamical models for Martian moons were typically established using analytical methods. The earliest analytical model was developed by Struve in 1911. He fitted observational data from the United States Naval Observatory collected between 1877 and 1909, calculating numerical values for the Martian moon's mean longitude, orbital elements, mean motion, and precession rate.
It was not until Shor that an analytical theory was introduced that incorporated the influence of tidal acceleration into the model~\cite{Shor1975TheMO}. This modification involved adding an acceleration term to the mean orbital longitude, and fitting was conducted using observational data spanning from 1877 to 1973.

With the continuous improvement in observational precision, analytical methods gradually revealed their limitations, leading to the brief emergence of semi-analytical approaches in dynamical modeling. The theory proposed by Chapront-Touzé~\cite{ChaprontTouz1988ESAPHOAS,ChaprontTouz1990OrbitsOT,ChaprontTouz1990PhobosPL} represents a high-precision semi-analytical model. It incorporates perturbations from the Sun and planets, the gravitational field of Mars, and mutual perturbations between Phobos and Deimos. This model is a more comprehensive semi-analytical approach to constructing dynamical models for Martian moons. Through fitting observational data from both ground-based and spacecraft observations spanning from 1877 to 1988, an ephemeris for the Martian moons was established.

With the development of deep space exploration, especially the increase in observational data and the continuous improvement in computing power, numerical methods, characterized by high computational accuracy and simplicity of formulas, have gradually replaced analytical methods as the primary approach in planetary orbit calculations. In 1977, Tolson~\cite{Tolson1977TheMO} used numerical methods and Viking spacecraft data to generate the first ephemeris for Phobos. In 2007, Lainey et al.~\cite{Lainey2007FirstNE} developed the first complete numerical integration ephemeris for the Martian moon. The dynamic model included a 10th-order non-spherical Martian gravity field~\cite{Lainey2004NewAE}, perturbations from the Sun, Jupiter, Saturn, Earth, and the Moon, mutual perturbations between the moons, tidal effects caused by the two moons on Mars, and the quadrupole moment of Phobos. Subsequently, Jacobson~\cite{Jacobson2010THEOA,Lainey2020MarsME} further incorporated the librational effects of the moons in the dynamical model, considering perturbations caused by the oscillation of the axis pointing toward Mars in the moon's body-fixed coordinate system within the orbital plane. Additionally, a new Mars rotation model was adopted, and the model was fitted with the Mars Express dataset, generating the Martian moon ephemeris MARS097.

In Jacobson's dynamical model, while the libration of the minimum moment of principal axis of Deimos in the orbital plane is considered, it does not incorporate the full rotation of the moon as is the case in the lunar dynamical model. 
Therefore, the aim of this paper is to establish a rotational model for Deimos in the inertial frame based on the Euler--Liouville rotational equations, following the approach used in the lunar dynamical model.
The purpose is to further develop the simple dynamical model for Martian moons and establish a full dynamical model for Deimos in the inertial frame. Simultaneously, based on the rotation model for Deimos, the study incorporates the influence of Deimos' own gravitational field when it orbits Mars, completing the full dynamical model of the rotational and orbital coupling for Deimos.

The first section of this paper will introduce the establishment process and distinctions between the simple model and the dynamical model considering the coupling of rotation and orbit. In the second section, the least squares method will be employed to introduce the data-fitting generation process and the adjustment model for precise orbit determination using the established dynamical model. Finally, a comparison will be made between the changes in coordinate axes during the orbital integration in inertial space for the \mbox{two models.}

\section{The Establishment of the Simple Model and Full Model}
\subsection{The Rotation of the Mars}

When describing the motion of Martian moons, the coordinate system chosen is the International Celestial Reference Frame (ICRF), while the gravitational field is represented in the Martian body-fixed coordinate system. Therefore, calculating the acceleration induced by the Martian gravity field on the moons requires an appropriate rotation model.
Konopliv et al. conducted an in-depth study of the Mars rotation based on data from the Pathfinder lander, establishing the most accurate current rotation model for Mars~\cite{Konopliv2016AnIJ}. According to this model, the method for converting positions from the Mars-fixed frame to the ICRF is as follows:
\vspace{-6pt}
\begin{equation}
	\label{eq7}
	\boldsymbol{R}_{I N}=R_Z(-N) R_X(-J) R_Z(-\psi) R_X(-I) R_Z(-\phi) R_Y(x_p) R_X(y_p)\boldsymbol{R}_{B F}
\end{equation}
where $\boldsymbol{R}_{I N}$ and $\boldsymbol{R}_{B F}$ represent the coordinates of Deimos in the inertial frame and the body-fixed frame, respectively.
$N$ is the angle in the plane of the Earth-mean-equator of J2000
(EME2000) from the vernal equinox (intersection of the mean ecliptic
and EME2000 planes) to the intersection of the Mars-mean-orbit of 
J2000 and EME2000 planes; $J$ is the inclination of the Mars-mean-orbit
plane relative to the EME2000 plane; $\psi$ is the angle in the 
Mars-mean-orbit plane from the intersection of the EME2000 and 
Mars-mean-orbit planes to the intersection of the Mars-mean-orbit and 
Mars-true-equator of date planes; $I$ is the inclination of the 
Mars true equator relative to the Mars-mean-orbit plane; $\phi$ is the 
angle in the plane of the Mars true equator from the intersection of 
the Mars-mean-orbit plane and Mars true equator to the Mars prime 
meridian; and $x_p$, $y_p$ represents the polar motion.

\subsection{Mars Gravity Field}

Due to the irregular shape and uneven mass distribution of Mars, when calculating the gravity on Deimos, it cannot be treated as an ideal homogeneous sphere. Non-spherical perturbations should be considered. The gravitational potential function is expanded into spherical harmonics as follows:
\begin{equation}
	\label{eq6}
	V=\frac{G M}{r} \sum_{n=2}^{\infty} \sum_{m=0}^n\left(\frac{R_{\text {mars }}}{r}\right)^n \bar{P}_{n m}(\sin \phi)\left[\bar{C}_{n m} \cos m \lambda+\bar{S}_{n m} \sin m \lambda\right]
\end{equation}

In this expression, $GM$ is the gravitational constant of Mars, $R_{\text {mars }}$ is the radius of Mars, $\bar{P}_{n m}$ represents the normalized associated Legendre function, and $\bar{C}_{n m}$ and $\bar{S}_{n m}$ are the normalized spherical harmonic coefficients of the Mars gravity field model. $\lambda$, $\phi$, and $r$ denote the longitude, latitude, and distance to the center of Mars in the fixed coordinate system of Deimos on Mars. When using the gravity field formula, it is necessary to follow the coordinate transformation method described in the previous section. First, the coordinates of Deimos in the inertial frame should be converted to the Mars-fixed frame. Subsequently, the integration calculations are performed by transforming from the Mars-fixed frame back to the inertial frame.

When computing gravitational field acceleration, a recursive method is commonly employed to avoid complex differentiation processes. For the Deimos orbit calculation, taking the example of a 10th-order Martian gravity field, the Cunningham~\cite{Cunningham1970OnTC} algorithm is used to calculate the gravitational field and its partial derivatives.

When calculating the Martian gravity potential, Cunningham provided a single recursion formula that can be used to calculate the gravity potential and the corresponding acceleration in Cartesian coordinates. The formula is defined as follows:

\begin{equation}
	\begin{aligned}
		& V_{n m}=\left(\frac{R}{r}\right)^{n+1} \cdot P_{n m}(\sin \phi) \cdot \cos m \lambda \\
		& W_{n m}=\left(\frac{R}{r}\right)^{n+1} \cdot P_{n m}(\sin \phi) \cdot \sin m \lambda
	\end{aligned}
\end{equation}
where the gravitational field can be represented as (non-normalized coefficients are used here):
\begin{equation}
	\text{U}=\frac{G M}{R} \sum_{n=0}^{\infty} \sum_{m=0}^n\left(C_{n m} V_{n m}+S_{n m} W_{n m}\right)
\end{equation}

Define the relationship:
\begin{equation}
	\begin{gathered}
		\left\{\begin{array}{l}
			V_{m m}=(2 m-1)\left(\frac{x R}{r^2} V_{m-1, m-1}-\frac{y R}{r^2} W_{m-1, m-1}\right) \\
			W_{m m}=(2 m-1)\left(\frac{x R}{r^2} W_{m-1, m-1}+\frac{y R}{r^2} V_{m-1, m-1}\right)
		\end{array}\right. \\
		\left\{\begin{array}{l}
			V_{n m}=\left(\frac{2 n-1}{n-m}\right) \cdot \frac{z R}{r^2} \cdot V_{n-1, m}-\left(\frac{n+m-1}{n-m}\right) \cdot \frac{R^2}{r^2} \cdot V_{n-2, m} \\
			W_{n m}=\left(\frac{2 n-1}{n-m}\right) \cdot \frac{z R}{r^2} \cdot W_{n-1, m}-\left(\frac{n+m-1}{n-m}\right) \cdot \frac{R^2}{r^2} \cdot V_{n-2, m}
		\end{array}\right.
	\end{gathered}
\end{equation}

Then, the acceleration of the gravitational field can be represented as: 

\begin{adjustwidth}{-\extralength}{0cm}
\begin{small}
\begin{equation}
	\begin{aligned}
	 \ddot{x}_{n m}  & =\left\{\begin{array}{c}
			\frac{G M}{R^2}\left(-C_{n 0} V_{n+1,1}\right), m=0 \\
			\frac{G M}{R^2} \cdot \frac{1}{2} \cdot\left\{-\left(C_{n m} V_{n+1, m+1}-S_{n m} W_{n+1, m+1}\right)+\frac{(n-m+2) !}{(n-m) !} \cdot\left(C_{n m} V_{n+1, m-1}+S_{n m} W_{n+1, m-1}\right)\right\}, m>0
		\end{array}\right.
	\end{aligned}
\end{equation}
\end{small}
\end{adjustwidth}

\begin{adjustwidth}{-\extralength}{0cm}
\begin{small}
\begin{equation}
	\begin{aligned}
		 \ddot{y}_{n m} & =\left\{\begin{array}{c}
			\frac{G M}{R^2}\left(-C_{n 0} W_{n+1,1}\right), m=0 \\
			\frac{G M}{R^2} \cdot \frac{1}{2} \cdot\left\{\left(-C_{n m} W_{n+1, m+1}+S_{n m} V_{n+1, m+1}\right)+\frac{(n-m+2) !}{(n-m) !} \cdot\left(-C_{n m} V_{n+1, m-1}+S_{n m} W_{n+1, m-1}\right)\right\}, m>0
		\end{array}\right.
	\end{aligned}
\end{equation}
\end{small}
\end{adjustwidth}

\begin{equation}
	\ddot{z}_{n m}=\frac{G M}{R^2} \cdot\left\{(n-m+1) \cdot\left(-C_{n m} V_{n+1, m}-S_{n m} V_{n+1, m}\right)\right\}
\end{equation}

\subsection{General Relativity Perturbation Model}

During the motion of Deimos, it is mainly influenced by the General Relativity effect generated by the mass of Mars. Since the eccentricity of Deimos' orbit is very small, it can be considered a nearly circular orbit. In this case, the acceleration induced by the General Relativity effect of Mars on Deimos is given by:
\begin{equation}
	\label{eq10}
	\boldsymbol{a}_{r e l}=-\frac{G M}{|\boldsymbol{r}|^3} \boldsymbol{r}\left(3 \frac{v^2}{c^2}\right)
\end{equation}
where $GM$ is the gravitational constant of Mars, 
$v$ is the velocity of Deimos in the Mars inertial frame, 
$\boldsymbol{r}$ is the position vector from the Mars to Deimos, and 
$c$ is the speed of light.

\subsection{Solid Tide Perturbation}

Due to the influence of the gravitational forces from other celestial bodies in the solar system, Mars experiences changes in its gravitational field, causing solid tidal effects. For a celestial body with a mass 
$M_j$ and a radius $a_j$, the solid tidal effect (according to elastic 
model pointed in~\cite{Yoder2003FluidCS} and also ~\cite{Konopliv2020DetectionOT}) on Deimos’
gravitational on Deimos' gravitational potential function is given by~\cite{Goossens2008LunarD2}:
\begin{equation}
	V_{\text {tide }}=\frac{k_2}{2} \frac{G M_j}{\left|\boldsymbol{r}_j\right|^3} \frac{a_j^5}{|\boldsymbol{r}|^3}\left(3\left(\boldsymbol{r}_j \cdot \boldsymbol{r}\right)^2-1\right)
\end{equation}
where $\boldsymbol{r}_j = \boldsymbol{r}-\boldsymbol{r}_m$ 
$\boldsymbol{r}_j$ is the position vector from the perturbing body to Deimos, 
$\boldsymbol{r}$ is the position vector from Mars to Deimos, 
$\boldsymbol{r}_m$ is the position vector from Mars to the perturbing body, and the Love number of Mars has a value of 0.169~\cite{Konopliv2020DetectionOT}. The solid tide model used in this study does not incorporate the tidal energy dissipation effects of Mars. We referred to previous studies on tidal energy  dissipation~\cite{Mignard1979TheEO,Efroimsky2007PhysicsOB,Efroimsky2012TIDALFA,Emelyanov2018InfluenceOT,Pou2022TidalCO}, and integrating the established dynamical model over ten years in numerical experiments reveals that tidal energy dissipation effects only cause differences of less than a few meters. Therefore, within the time scale of this study (10 years), we have not considered the influence of \mbox{tidal dissipation.}

\subsection{Simple Model}
The current Deimos dynamical model is referred to as a simplified model. In the simple Deimos dynamical model, the librations of Deimos are considered. Jacobson~\cite{Jacobson2010THEOA} introduced the specific expression for this effect:
\begin{equation}
	\boldsymbol{F}=-\frac{3}{2} \mu\left(\frac{R^2}{r^4}\right)\left[\left(J_2+6 C_{22} \cos 2 \theta\right) \hat{\boldsymbol{r}}+4 C_{22} \sin 2 \theta \hat{\boldsymbol{t}}\right]
\end{equation}
\begin{equation}
	\theta=f-M+\frac{2 e}{1-\frac{C}{3(B-A)}} \sin M
\end{equation}
where $\mu$ represents the gravitational constant of Mars; $R$ is the radius of Deimos; $r$ represents the distance from Mars to Deimos; 
$J_2$ is the second-degree sectoral harmonic coefficient of Deimos; 
$C_{22}$ is the second-degree tesseral harmonic coefficient of Deimos; 
$f$ and $M$ are the true anomaly and mean anomaly of Deimos' orbit, respectively; 
$e$ is the eccentricity of the orbit; 
$A$, $B$, and $C$ are the moments of inertia with respect to the principal axes  of Deimos;  
$\hat{\boldsymbol{r}}$ is the unit vector from Mars to Deimos; and 
$\hat{\boldsymbol{t}}$ is the unit vector in the Deimos orbital plane that is perpendicular to the velocity vector of Deimos. This model is the dynamical model used in the current ephemeris.

\subsection{Full Model}
The main difference between the full model and the simple model lies in the consideration of the complete rotation of Deimos in inertial space in the full model. There have been some research efforts on the rotation of Phobos~\cite{Rambaux2012RotationalMO,Yang2020AnEM}. In this section, we will establish the rotation model of Deimos based on the theory of rigid body rotation. Based on the established rotation model, it is possible to calculate the gravitational force of Deimos on Mars and the acceleration of Deimos caused by the reaction force it experiences due to the gravitational interaction with Mars. Thus, a dynamical model for the coupled rotation and orbit of Deimos has been established. The gravitational field coefficients for Deimos are listed in Table~\ref{tab1}~\cite{Rubincam1995TheGF}. This study further advances the current dynamical model that only considers the simple librations of Deimos. It comprehensively calculates the rotation of Deimos in inertial space due to torques from celestial bodies such as Mars and the Sun.
\begin{table}[H] 
	\caption{Normalized gravitational coefficients of Deimos in units of $10^{-2}$.\label{tab1}}
	\newcolumntype{C}{>{\centering\arraybackslash}X}
	\begin{tabularx}{\textwidth}{CCC}
		\toprule
		\boldmath{$(l,m)$} & \boldmath{$C_{n m} (\times 10^{-2})$} & \boldmath{$S_{n m} (\times 10^{-2})$}\\
		\midrule
		(1, 0)		& 0			& 0\\
		(1, 1)		& 0			& 0\\
		(2, 0)		& $-$4.827			& 0\\
		(2, 1)		& $-$0.0727			& 0.362\\
		(2, 2)		& 4.773			& 0.122\\
		(3, 0)		& 0.970			& 0\\
		(3, 1)		& 1.430			& 0.0521\\
		(3, 2)		& $-$1.108			& $-$0.499\\
		(3, 3)		& $-$0.121			& 0.728\\
		\bottomrule
	\end{tabularx}
\end{table}
The differential equations governing the rotational motion of a rigid body are classically represented by the Euler--Liouville equations, and their form in an inertial frame is \mbox{as follows:}

\begin{equation}
	\label{eq13}
	\frac{d(I \boldsymbol{\omega})}{d t}+\boldsymbol{\omega} \times I \boldsymbol{\omega}=\boldsymbol{N}
\end{equation}
where $N$ is the torque acting on Deimos, $\boldsymbol{\omega}$ represents the rotation of Deimos, and $I$ is inertia tensor matrix.

When describing the rotation of Deimos, we take the center of mass of Deimos as the origin, the three principal axes of Deimos as the coordinate axes, and establish the Deimos Principal Axis (PA) reference frame rotating with Deimos. Simultaneously, we can obtain the celestial reference frame of the Deimos center of mass with coordinates aligned with ICRS. Assuming Deimos as an ideal rigid body, the transformation between the Deimos body-fixed coordinate system and the inertial frame is determined by three Euler angles ($\phi$, $\theta$, $\psi$) describing the rotation of the rigid body. This transformation process is expressed as:
\begin{equation}
	\boldsymbol{R}_{I N}=R_Z(-\phi) R_X(-\theta) R_Z(-\psi) \boldsymbol{R}_{B F}
\end{equation}

$\boldsymbol{\omega}$ represents the rotation of the rigid-body moon under the action of torque in the inertial coordinate system, and the relationship between $\boldsymbol{\omega}$ and the three Euler angles and the rates of change of the three Euler angles can be obtained:
\begin{equation}
	\begin{gathered}
		\omega_x=\dot{\phi} \sin \theta \sin \psi+\dot{\theta} \cos \psi \\
		\omega_y=\dot{\phi} \sin \theta \cos \psi-\dot{\theta} \sin \psi \\
		\omega_z=\dot{\phi} \cos \theta+\dot{\psi}
	\end{gathered}
\end{equation}

Solving for the first derivative of $\boldsymbol{\omega}$, we can obtain the accelerations of the three \mbox{Euler angles:}

\begin{equation}
	\label{eq16}
	\begin{gathered}
		\ddot{\phi}=\frac{\dot{\omega}_x \sin \psi+\dot{\omega}_x \cos \psi+\dot{\theta}(\dot{\psi}-\dot{\phi} \cos \theta)}{\sin \theta} \\
		\ddot{\theta}=\dot{\omega}_x \cos \psi-\dot{\omega}_y \sin \psi-\dot{\phi} \dot{\psi} \sin \theta \\
		\ddot{\psi}=\dot{\omega}_z-\ddot{\phi} \cos \theta+\dot{\phi} \psi \sin \theta
	\end{gathered}
\end{equation}

However, when employing Equation~\eqref{eq16}, one may encounter singular situations. In such cases, using Wisdom angles instead of Euler angles might be more beneficial~\cite{Wisdom1984TheCR}.
According to the Euler--Liouville equation, the value of $\dot{\omega}$ depends on the torque acting on the Deimos and the moment of inertia. Taking the first derivative of Equation~\eqref{eq13} yields \mbox{the expression:}
\begin{equation}
	\dot{\boldsymbol{\omega}}=I^{-1}(\boldsymbol{N}-\boldsymbol{\omega} \times I \boldsymbol{\omega})
\end{equation}

$I$ is the inertia tensor matrix, which can be expressed as:
\begin{equation}
	I=\left(\begin{array}{lll}
		A & 0 & 0 \\
		0 & B & 0 \\
		0 & 0 & C
	\end{array}\right)
\end{equation}

The torque acting on Deimos during its rotation can be divided into three components: (1) the torque produced by Mars as a point mass in the Deimos gravitational field; (2) the torque exerted by the Sun, treating it as a point mass due to the presence of the Deimos gravitational field; and (3) the torque generated in the Deimos due to the irregular shape of Mars, under the influence of the Deimos gravitational field~\cite{Pavlov2016DeterminingPO}. The formulas can be expressed as:
\begin{equation}
	\boldsymbol{N}=\boldsymbol{N}_{\oplus}+\boldsymbol{N}_{\odot}+\boldsymbol{N}_{\oplus \mathbf{F}}
\end{equation}
where $\boldsymbol{N}_{\oplus}$ and $\boldsymbol{N}_{\odot}$ represent the torques generated by the point masses Mars and the Sun, and $\boldsymbol{N}_{\oplus \mathbf{F}}$ represents the torque produced by the shape effect of Mars ~\cite{Breedlove1977ANS}. $\boldsymbol{N}_{\oplus \mathbf{F}}$ is generated by the second-order term of the Martian gravitational field and is mainly related to the value of Mars. In numerical experiments, the torque generated by Phobos is a magnitude smaller than that of Mars and the Sun. Therefore, the model does not consider the impact of the torque generated by Phobos. The force exerted by Mars in the moon's gravitational field is expressed as:
\begin{equation}
	\boldsymbol{F}=\mathrm{M}_{\oplus}\left(-\nabla \mathrm{U}_{\mathrm{d}}\right)
\end{equation}

The torque produced by this force in the Deimos is expressed as:
\begin{equation}
	\boldsymbol{N}_{\oplus}=\mathrm{M}_{\oplus}\left(\boldsymbol{r}_{\mathrm{d}} \times \nabla \mathrm{U}_{\mathrm{d}}\right)
\end{equation}

$\mathrm{M}_{\oplus}$ represents the mass of Mars, $\boldsymbol{r}_{\mathrm{d}}$ represents the coordinates of Deimos in the Mars-centered inertial frame, and $\mathrm{U}_{\mathrm{d}}$ represents the gravity field of Deimos. And $\boldsymbol{N}_{\oplus \mathbf{F}}$ represents the torque induced by the oblateness of Mars and the shape effect of Deimos. The expression is given by 
~\cite{Pavlov2016DeterminingPO}:
\vspace{-6pt}
\begin{adjustwidth}{-\extralength}{0cm}
\begin{small}
\begin{equation}
	\begin{aligned}
		 \frac{\mathbf{N}_{\oplus \mathbf{F}}}{m_D}=\frac{15 \mu_M R_M^2 J_{2 M}}{2 r_{M D}^5} {\left[\left(1-7\left(\hat{\boldsymbol{r}}_{M D} \cdot \widehat{\boldsymbol{p}}\right)\right)^2\right)\left(\hat{\boldsymbol{r}}_{M D} \times \frac{I}{m_D} \hat{\boldsymbol{r}}_{M D}\right)+2\left(\hat{\boldsymbol{r}}_{M D} \cdot \widehat{\boldsymbol{p}}\right)\left(\hat{\boldsymbol{r}}_{M D} \times \frac{I}{m_D} \widehat{\boldsymbol{p}}+\widehat{\boldsymbol{p}} \times \frac{I}{m_D} \hat{\boldsymbol{r}}_{M D}\right) } \left.-\frac{2}{5}\left(\hat{\boldsymbol{p}} \times \frac{I}{m_D} \widehat{\boldsymbol{p}}\right)\right]
	\end{aligned}
\end{equation}
\end{small}
\end{adjustwidth}

$m_D$ is the mass of Deimos, $J_{2 M}$ represents the value of the Mars flattening, $r_{M D}$ is the distance between Deimos and Mars, $\hat{\boldsymbol{r}}$ is the unit vector pointing from Deimos to Mars, $\widehat{\boldsymbol{p}}$ is the direction of Mars's polar axis, and all vectors are described in the Deimos body-fixed coordinate system.

\begin{figure}[H]
	\includegraphics[width=10.5 cm]{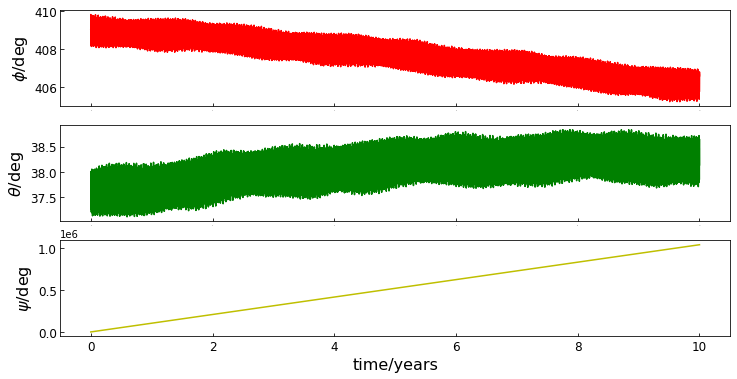}
	\caption{The change of Euler angles with respect to the integration time.}
	\label{fig1}
\end{figure}

\begin{figure}[H]
	\includegraphics[width=10.5 cm]{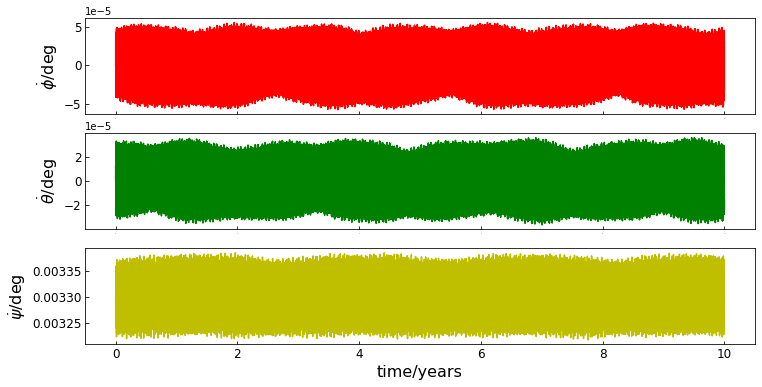}
	\caption{The change of Euler angle rates with respect to the integration time.}
	\label{fig2}
\end{figure}

According to the established model, the initial Euler angles of Deimos at the J2000 epoch are provided by Brent et al.  \cite{Archinal2018ReportOT}. The integration of the established rotational differential equations using the ABM integration algorithm yields the results for the three Euler angles as shown in Figures~\ref{fig1} and \ref{fig2}.
\unskip

\subsection{Result of the Model}

\begin{figure}[H]
	\includegraphics[width=10.5 cm]{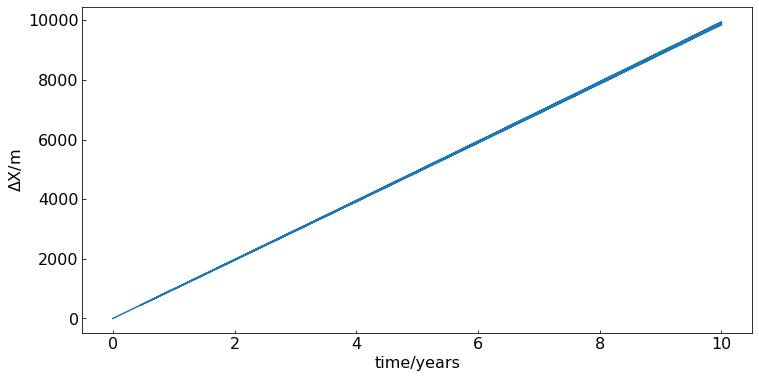}
	\caption{The difference in position between the Deimos full dynamics model and simple model.}
	\label{fig3}
\end{figure}

At this point, we have considered factors influencing the motion of Deimos, utilizing a second-order gravity field for Deimos. We have established a dynamical model for the coupled rotation and orbit of Deimos. The initial value of Deimos' position at the J2000 epoch is provided by MARS097. The integration orbit for both the simple dynamical model and the full model with rotation--orbit coupling established in this study are compared in Figure~\ref{fig3}.

The experimental results indicate that, considering the J2000 epoch as the integration starting point with a time span of 10 years, the difference in the computed results between the two dynamical models is around 10 km. The difference increases linearly over time, primarily due to the inclusion of rotation and the gravitational field of Deimos itself in the dynamical model. This discrepancy suggests that considering the complete rotation effect of Deimos in the dynamical model is essential for improving its accuracy.

\begin{figure}[H]
	\includegraphics[width=10.5 cm]{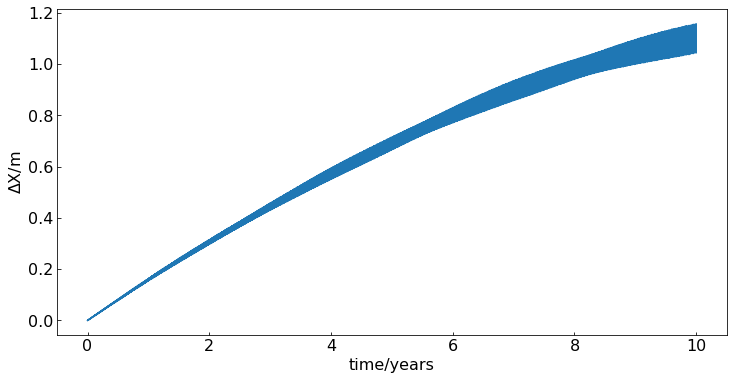}
	\caption{The difference in position in using the second-order gravity field and the third-order gravity field in the full model for Deimos.}
	\label{fig4}
\end{figure}

When considering the third-order gravity field of Deimos, the integrated orbit of Deimos shows differences compared to the integrated orbit of the dynamical model considering only the second-order gravity field of Deimos as shown in Figure~\ref{fig4}.
The model results indicate that, after integrating for 10 years, the difference between the second- and third-order gravity fields of Deimos is only on the order of meters, which is an extremely small value. Therefore, when constructing a comprehensive model for Deimos, considering gravity field orders up to the second order is sufficient.


\section{Adjustment Model}

In this section, we conducted a precise orbit determination for the established full dynamical model of Deimos, based on the methodology employed in the precise orbit determination of artificial satellites \cite{Beletskiy1959TheLO}. When selecting data for the fitting process, we initially utilized observations from the ephemeris MARS097 as the basis for refining the simple dynamical model. The resulting fitted orbit serves as the input data for the full model. This approach ensures consistency in the parameters between the two dynamical models, with the integrated results highlighting differences arising solely from distinct rotational modes between the simple and full models.

\subsection{Variation Equations}

For a specified epoch, changes in the initial values of the state vector result in variations in the position and velocity at subsequent epochs, represented by the state transition matrix. The variation equations can be established from the state transition matrix:
\begin{equation}
	\frac{d(\Phi, S)}{d x}=\left(\begin{array}{cc}
		0_{3 \times 3} & 1_{3 \times 3} \\
		\frac{\partial \boldsymbol{a}}{\partial \boldsymbol{r}} & \frac{\partial \boldsymbol{a}}{\partial \boldsymbol{v}}
	\end{array}\right)_{6 \times 6} \cdot(\Phi, S)+\left(\begin{array}{cc}
		0_{3 \times 6} & 0_{3 \times n_p} \\
		0_{3 \times 6} & \frac{\partial \boldsymbol{a}}{\partial p}
	\end{array}\right)_{6 \times\left(6+n_p\right)}
\end{equation}

For the full model, involving both the rotation and orbital parameters (12 state variables in total), the structure of the state transition matrix is outlined in Equation~\eqref{ta}:

\begin{equation}
	\label{ta}
	\begin{array}{|c|c|c|c|c|c|c|c|c|c|c|c|c|}
		\hline & x & y & z & v_x & v_y & v_z & \phi & \theta & \psi & \dot{\phi} & \dot{\theta} & \dot{\psi} \\
		\hline v_x & 0 & 0 & 0 & 1 & 0 & 0 & 0 & 0 & 0 & 0 & 0 & 0 \\
		\hline v_y & 0 & 0 & 0 & 0 & 1 & 0 & 0 & 0 & 0 & 0 & 0 & 0 \\
		\hline v_z & 0 & 0 & 0 & 0 & 0 & 1 & 0 & 0 & 0 & 0 & 0 & 0 \\
		\hline a_x & \frac{\partial a_x}{\partial x} & \frac{\partial a_x}{\partial y} & \frac{\partial a_x}{\partial z} & \frac{\partial a_x}{\partial v_x} & \frac{\partial a_x}{\partial v_y} & \frac{\partial a_x}{\partial v_z} & \frac{\partial a_x}{\partial \phi} & \frac{\partial a_x}{\partial \theta} & \frac{\partial a_x}{\partial \psi} & 0 & 0 & 0 \\
		\hline a_y & \frac{\partial a_y}{\partial x} & \frac{\partial a_y}{\partial y} & \frac{\partial a_y}{\partial z} & \frac{\partial a_y}{\partial v_x} & \frac{\partial a_y}{\partial v_y} & \frac{\partial a_y}{\partial v_z} & \frac{\partial a_y}{\partial \dot{\phi}} & \frac{\partial a_y}{\partial \theta} & \frac{\partial a_y}{\partial \psi} & 0 & 0 & 0 \\
		\hline a_z & \frac{\partial a_z}{\partial x} & \frac{\partial a_z}{\partial y} & \frac{\partial a_z}{\partial z} & \frac{\partial a_z}{\partial v_x} & \frac{\partial a_z}{\partial v_y} & \frac{\partial a_z}{\partial v_z} & \frac{\partial a_z}{\partial \phi} & \frac{\partial a_z}{\partial \theta} & \frac{\partial a_z}{\partial \psi} & 0 & 0 & 0 \\
		\hline \dot{\phi} & 0 & 0 & 0 & 0 & 0 & 0 & 0 & 0 & 0 & 1 & 0 & 0 \\
		\hline \dot{\theta} & 0 & 0 & 0 & 0 & 0 & 0 & 0 & 0 & 0 & 0 & 1 & 0 \\
		\hline \dot{\psi} & 0 & 0 & 0 & 0 & 0 & 0 & 0 & 0 & 0 & 0 & 0 & 1 \\
		\hline \ddot{\phi} & \frac{\partial \ddot{\phi}}{\partial x} & \frac{\partial \ddot{\phi}}{\partial y} & \frac{\partial \ddot{\phi}}{\partial z} & 0 & 0 & 0 & \frac{\partial \ddot{\phi}}{\partial \dot{\phi}} & \frac{\partial \ddot{\phi}}{\partial \theta} & \frac{\partial \ddot{\phi}}{\partial \psi} & \frac{\partial \ddot{\phi}}{\partial \dot{\phi}} & \frac{\partial \ddot{\phi}}{\partial \dot{\theta}} & \frac{\partial \ddot{\phi}}{\partial \dot{\psi}} \\
		\hline \ddot{\theta} & \frac{\partial \ddot{\theta}}{\partial x} & \frac{\partial \ddot{\theta}}{\partial y} & \frac{\partial \ddot{\theta}}{\partial z} & 0 & 0 & 0 & \frac{\partial \ddot{\theta}}{\partial \phi} & \frac{\partial \ddot{\theta}}{\partial \theta} & \frac{\partial \ddot{\theta}}{\partial \psi} & \frac{\partial \ddot{\theta}}{\partial \dot{\phi}} & \frac{\partial \ddot{\theta}}{\partial \dot{\theta}} & \frac{\partial \ddot{\theta}}{\partial \dot{\psi}} \\
		\hline \ddot{\psi} & \frac{\partial \ddot{\psi}}{\partial x} & \frac{\partial \ddot{\psi}}{\partial y} & \frac{\partial \ddot{\psi}}{\partial z} & 0 & 0 & 0 & \frac{\partial \ddot{\psi}}{\partial \phi} & \frac{\partial \ddot{\psi}}{\partial \theta} & \frac{\partial \ddot{\psi}}{\partial \psi} & \frac{\partial \ddot{\psi}}{\partial \dot{\phi}} & \frac{\partial \ddot{\psi}}{\partial \dot{\theta}} & \frac{\partial \ddot{\psi}}{\partial \dot{\psi}} \\
		\hline
	\end{array}
\end{equation}

Let	$\ddot{\varphi}\equiv F_1, \ddot{\theta}\equiv F_2 , \ddot{\psi}\equiv F_3$, and calculate the first derivative of $F_1, F_2, F_3$:
\begin{equation}
	\begin{gathered}
		\frac{\partial F_1}{\partial \phi}=\csc \theta\left(\sin \psi \frac{\partial \dot{\omega}_x}{\partial \phi}+\cos \psi \frac{\partial \dot{\omega}_y}{\partial \phi}\right) \\
		\frac{\partial F_2}{\partial \phi}=\cos \psi \frac{\partial \dot{\omega}_x}{\partial \phi}+\sin \psi \frac{\partial \dot{\omega}_y}{\partial \phi} \\
		\frac{\partial F_3}{\partial \phi}=\frac{\partial \dot{\omega}_z}{\partial \phi}-\cos \theta \frac{\partial F_1}{\partial \phi}
	\end{gathered}
\end{equation}

\begin{equation}
	\begin{gathered}
		\frac{\partial F_1}{\partial \theta}=\csc \theta\left(\sin \psi \frac{\partial \dot{\omega}_x}{\partial \theta}+\cos \psi \frac{\partial \dot{\omega}_y}{\partial \theta}\right)-\\\cot \theta \csc \theta\left(\dot{\omega}_x \sin \psi+\dot{\omega}_y \cos \psi\right)-\dot{\theta} \dot{\psi} \cot \theta \csc \theta+\dot{\theta} \dot{\phi} \cot \theta \csc ^2 \theta \\
		\frac{\partial F_2}{\partial \theta}=\cos \psi \frac{\partial \dot{\omega}_x}{\partial \theta}-\sin \psi \frac{\partial \dot{\omega}_y}{\partial \theta}-\dot{\theta} \dot{\phi} \cos \theta \\
		\frac{\partial F_3}{\partial \theta}=\frac{\partial \dot{\omega}_z}{\partial \theta}-\cos \psi \frac{\partial F_1}{\partial \theta}+\sin \theta F_1+\dot{\theta} \dot{\phi} \cos \theta
	\end{gathered}
\end{equation}

\begin{equation}
	\begin{gathered}
		\frac{\partial F_1}{\partial \psi}=\csc \theta\left(\dot{\omega}_x \cos \psi+\sin \psi \frac{\partial \dot{\omega}_x}{\partial \psi}-\dot{\omega}_y \sin \psi+\cos \psi \frac{\partial \dot{\omega}_y}{\partial \psi}\right) \\
		\frac{\partial F_2}{\partial \psi}=\cos \psi \frac{\partial \dot{\omega}_x}{\partial \psi}-\sin \psi \dot{\omega}_x-\cos \psi \dot{\omega}_y-\sin \psi \frac{\partial \dot{\omega}_y}{\partial \psi} \\
		\frac{\partial F_3}{\partial \psi}=\frac{\partial \dot{\omega}_z}{\partial \psi}-\cos \theta \frac{\partial F_1}{\partial \psi}
	\end{gathered}
\end{equation}

According to this method, the expression for the first derivative with respect to Euler angular rates $\frac{\partial F}{\partial \dot{A}}$ can also be derived. According to the Euler--Liouville equation, the expression for the partial derivative of $\dot{\boldsymbol{\omega}}$ with respect to Euler angles ($\frac{\partial \dot{\boldsymbol{\omega}}}{\partial A}$) is given by:
\begin{equation}
	\frac{\partial \dot{\boldsymbol{\omega}}}{\partial A}=I^{-1}\left(\frac{\partial \boldsymbol{N}}{\partial A}-\frac{\partial \boldsymbol{\omega}}{\partial A} \times I \boldsymbol{\omega}-\boldsymbol{\omega} \times I \frac{\partial \boldsymbol{\omega}}{\partial A}\right)
\end{equation}
\begin{equation}
	\frac{\partial \boldsymbol{N}}{\partial A}=M_b\left(\frac{\partial \boldsymbol{r_b}}{\partial A} \times \nabla U_b+\boldsymbol{r_b} \times \frac{\partial \nabla U_b}{\partial A}\right)
\end{equation}

Coordinates in the body-fixed coordinate system of Deimos are denoted by $(\xi_1, \xi_2, \xi_3)$. According to the transformation matrix from inertial frame to body-fixed frame, the expression for $\frac{\partial r}{\partial A}$ can be solved as:
\begin{equation}
	\begin{gathered}
		\frac{\partial x}{\partial \phi}=(-\cos \phi \sin \psi-\sin \psi \cos \theta \cos \phi) \xi_1+(\cos \phi \cos \psi-\sin \psi \cos \theta \sin \phi) \xi_2 \\
		\frac{\partial x}{\partial \theta}=\sin \psi \sin \theta \sin \phi \xi_1-\sin \psi \sin \theta \cos \phi \xi_2+\sin \psi \sin \theta \xi_3 \\
		\frac{\partial x}{\partial \psi}=(-\cos \phi \sin \psi-\sin \psi \cos \theta \cos \psi) \xi_1+(-\sin \phi \sin \psi+\cos \psi \cos \theta \sin \phi) \xi_2+\\\cos \psi \cos \theta \xi_3
	\end{gathered}
\end{equation}
\begin{equation}
	\begin{gathered}
		\frac{\partial y}{\partial \phi}=(-\cos \psi \sin \phi-\cos \psi \cos \theta \cos \phi) \xi_1+(\cos \phi \sin \psi-\sin \phi \cos \theta \cos \psi) \xi_3 \\
		\frac{\partial y}{\partial \theta}=\cos \psi \sin \theta \sin \phi \xi_1-\cos \psi \sin \theta \cos \phi \xi_2+\cos \psi \cos \theta \xi_3 \\
		\frac{\partial y}{\partial \psi}=(-\cos \phi \cos \psi+\sin \psi \cos \theta \sin \phi) \xi_1+(-\sin \phi \cos \psi+\sin \psi \cos \theta \cos \phi) \xi_2-\\\sin \psi \sin \theta \xi_3
	\end{gathered}
\end{equation}

\begin{equation}
	\begin{gathered}
		\frac{\partial z}{\partial \phi}=\sin \psi \cos \phi \xi_1+\sin \theta \sin \phi \xi_3 \\
		\frac{\partial z}{\partial \theta}=\cos \theta \sin \phi \xi_1-\cos \theta \cos \phi \xi_2-\sin \theta \xi_3 \\
		\frac{\partial z}{\partial \psi}=0
	\end{gathered}
\end{equation}

\begin{figure}[H]
	\includegraphics[width=10.5 cm]{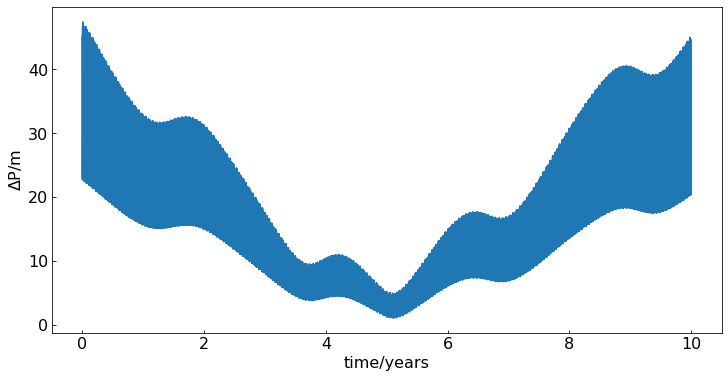}
	\caption{The discrepancy between the precise orbit determination of the full dynamics model and the position of MARS097.}
	\label{fig5}
\end{figure}

\begin{figure}[H]
	\includegraphics[width=10.5 cm]{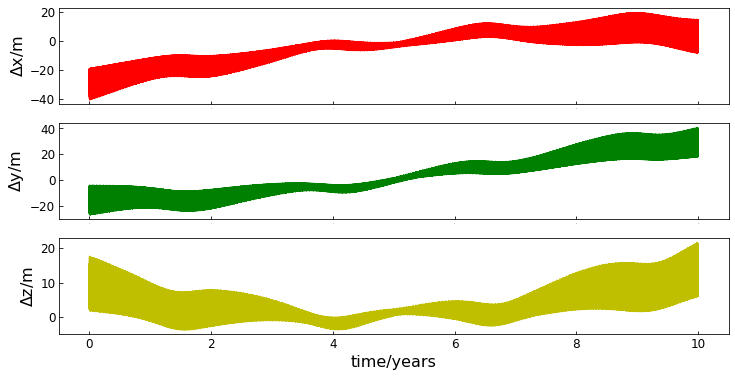}
	\caption{The discrepancy in XYZ coordinates between the precise orbit determination of the full dynamics model and the position of MARS097.}
	\label{fig6}
\end{figure}

Finally, employing the precision orbit determination method, the computed state transition matrix is synchronously integrated with the orbit. Subsequently, utilizing a least-squares fit to observational data, the calculated results are compared with the MARS097 ephemeris as depicted in Figures~\ref{fig5} and~\ref{fig6}.
\unskip
With the J2000 epoch as the integration starting point, the maximum discrepancy between the final results of the data fitting using the full dynamical model for Deimos and the MARS097 ephemeris is approximately 50 m. This demonstrates the stability and reliability of the established dynamical model and adjustment approach.

\subsection{Variation of the Deimos Principal Axes in the Inertial Frame}

\begin{figure}[H]
	\includegraphics[width=10.5 cm]{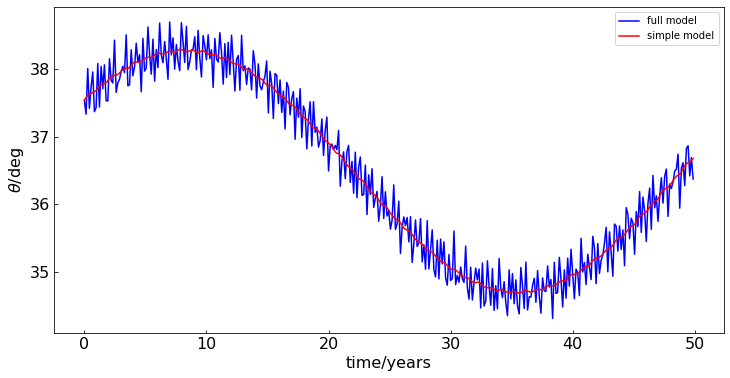}
	\caption{The orientation of the pole for Deimos in the inertial frame in both the simple model and the full model.}
	\label{fig7}
\end{figure}

The fundamental distinction between the full model and the simple dynamical model lies in the incorporation of the Deimos's full rotation in the inertial frame. In the simple model, the moon's polar axis direction is solely perpendicular to the orbital plane. Nevertheless, when accounting for the complete rotation, the moon's polar axis undergoes oscillation in the inertial frame as the moon rotates. Figure~\ref{fig7} provides a comparison of the directional deviation of the polar axis under the two motion modes.

\begin{figure}[H]
	\includegraphics[width=10.5 cm]{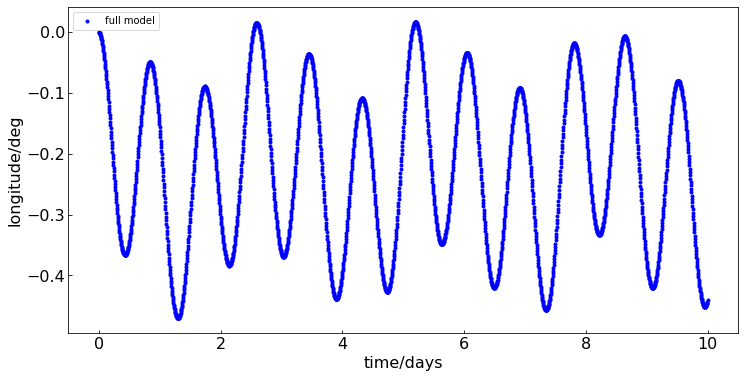}
	\caption{The angle between the minimum moment of inertia axis and the Mars position vector in the longitude direction in the complete model.}
	\label{fig8}
\end{figure}

\begin{figure}[H]
	\includegraphics[width=10.5 cm]{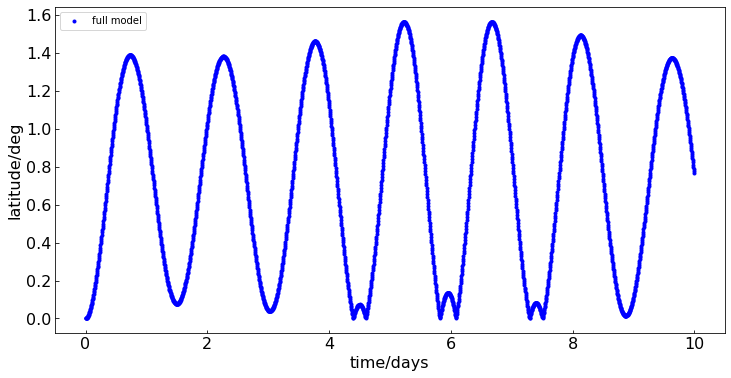}
	\caption{The angle between the minimum moment of inertia axis and the Mars position vector in the latitude direction in the full model.}
	\label{fig9}
\end{figure}

Starting from the J2000 epoch, Figure~\ref{fig7} illustrates the evolution of the angle between the polar axis and the vector (0, 0, 1) in the inertial frame over a 50-year period for both motion models. In the simplified model, the polar axis exhibits relative stability in space, primarily influenced by the precession effect. However, in the full model, accounting for the rotational forces from Mars and the Sun, the polar axis undergoes intricate motion in the inertial frame. Figures~\ref{fig8} and~\ref{fig9} represent the angles between the minimum moment of inertia axis of the moon in the moon-fixed coordinate system and the Mars position vector in the longitude and latitude directions, respectively. The results reveal that the moon's amplitude in the longitude direction is approximately 0.5 degrees, and in the latitude direction, it does not exceed 2 degrees.
\unskip

\section{Conclusions}

This study presents a novel model describing the motion of Deimos in an inertial frame. Building upon existing models, it thoroughly calculates the rotational effects induced by the irregular shape of Deimos itself and the torques exerted by large celestial bodies (Mars, Sun) in the solar system in inertial space. Furthermore, it considers the coupling effect resulting from rotation and orbit, establishing a full dynamical model for the rotation of Deimos coupled with its orbit. The model is integrated using the Adams--Bashforth--Moulton (ABM) integration algorithm to concurrently solve the 12 elements governing rotation and orbital dynamics in the differential equations.

To validate the model, precise orbit determination techniques for artificial satellites are employed to fit data to the full model. The results demonstrate the stability and reliability of the established model. Additionally, the study derives an analytical expression for the solution of the coupling effect between Deimos' rotation and orbit. The variational equation, starting from the J2000 epoch, is integrated over ten years. The final outcomes reveal differences of less than 50 m when compared to the current Deimos ephemeris MARS097, showcasing its practical applicability in engineering.

As various exploration missions targeting the Martian moons continue in the future, it is anticipated that a considerable amount of high-precision data on these moons will be generated. Leveraging these data in conjunction with the full dynamical model established in this study, there is potential to create a new generation of more accurate ephemerides for the Martian moons. Additionally, by utilizing the full model established in this paper along with future observational data, there is the prospect of quantitatively studying orbital resonances among celestial bodies in the solar system~\cite{Peale1976OrbitalRI,Goldreich1966SpinorbitCI} and the Lense--Thirring effect~\cite{Ciufolini2004ACO,Iorio2007OnTL,Iorio2008JunoTA} in orbital mechanics. This, in turn, holds the potential to further enhance the accuracy of the Deimos' dynamical model.

The model established in this paper lays the foundation for further constructing an accurate ephemeris for Deimos. It demonstrates versatility in solving the dynamical equations for celestial bodies similar to Deimos and can serve as a reference for developing numerical ephemerides for major celestial bodies in the solar system.

\vspace{6pt} 




\authorcontributions{This paper is a collaborative work by all of the authors. K.H.: Conceptualization, methodology, software, and writing---original draft preparation; L.Z.: software; Y.Y.: Conceptualization, methodology; M.Y.: methodology, software; Y.L.: Conceptualization. All authors have read and agreed to the published version of the manuscript.}

\funding{This research is supported by the National Natural Science Foundation of China (12033009, 12103087), the National Key Research and Development Program of China (2021YFA0715101), the International Partnership Program of Chinese Academy of Sciences (020GJHZ2022034FN), the Yunnan Province Foundation (202201AU070225, 202301AT070328, 202401AT070141), the Yunnan Revitalization Talent Support Program, grant from Key Laboratory of Lunar and Deep Space Exploration, Chinese Academy of Sciences (LDSE202004), open research fund of state key laboratory of information engineering in surveying, mapping and remote sensing, Wuhan University (21P02), and supported by Key Laboratory of TianQin Project (Sun Yat-sen University), Ministry of Education. M. Ye is supported by Natural Science Foundation of Hubei Province, China, No. 2022CFB123, Space Optoelectronic Measurement and Perception Laboratory, Beijing Institute of Control Engineering (No. LabSOMP-2023-09).}



\dataavailability{ The data presented in this study are available on request from the corresponding author. The data are not publicly available due to privacy.} 


\acknowledgments{Deimos' ephemerides files can be downloaded from 
 \url{https://naif.jpl.nasa.gov/pub/naif/INSIGHT/kernels/spk/mar097s.bsp/} (accessed on 1 May 2023)} (Mars097), and \url{ftp://ftp.imcce.fr/pub/ephem/NOE/} (accessed on 1 May 2023) (NOE-4-2020). And the EPM ephemeris is obtained from the Online Ephemeris Service \url{http://iaaras.ru/en/dept/ephemeris/online/} (accessed on 1 May 2023).

\conflictsofinterest{The authors declare no conflicts of interest. 
} 



\begin{adjustwidth}{-\extralength}{0cm}

\reftitle{References}

\PublishersNote{}
\end{adjustwidth}
\end{document}